\documentclass[aps,prb,amsmath]{revtex4-2}

\usepackage{amsmath}
\usepackage[mathlines]{lineno}%
\usepackage{dsfont}
\usepackage{mathtools}
\usepackage{caption}
\usepackage{graphicx}
\usepackage{epstopdf}
\usepackage{dcolumn}
\usepackage{bm}
\usepackage{color}
\usepackage{subfig}
\usepackage{float}
\usepackage[normalem]{ulem}
\usepackage{natbib}

\newcommand{\modsubfigimg}[3][,]{%
	\setbox1=\hbox{\includegraphics[#1]{#3}}
	\leavevmode\rlap{\usebox1}
	\rlap{\hspace*{155pt}\raisebox{\dimexpr\ht1-2\baselineskip}{#2}}
	\phantom{\usebox1}
}

\begin{document}
	
\title{Erratum: Caroli formula in near-field heat transfer between\\parallel graphene sheets [PHYSICAL REVIEW B 96, 155437 (2017)]}
	
\author{Jia-Huei Jiang}
\author{Jian-Sheng Wang}

\date{\today}


\maketitle

This erratum aims to correct 1) the wrong expressions, 2) some typographical errors,\,3) some erroneous points made in discussion of the disparity of heat flux ratios between our full RPA model and the local conductivity model, and 4) the lower bound of the characteristic distance scale comparable to the graphene thermal length, appearing in the original paper \cite{Caroli_PhiG}, which might confuse or mislead readers. Collectively, they don't change the key physics we sought to present. In the following, whenever we say the/ our original paper, it refers to Ref. \cite{Caroli_PhiG}.\\
First, we would like to correct wrong expressions:
\begin{enumerate}
	\item In Eq. (A1), the electronic annihilation and creation operators were given wrongly in the paper. They should be as follows:
	$c_{\textbf{k}}^{(l)}=\left(c_{A\,\textbf{k}}^{\quad (l)}, c_{B\,\textbf{k}}^{\quad (l)}\right)^{T}$; ${c_{\textbf{k}}}^{\dagger\,(l)}=\left({c_{A\,\textbf{k}}}^{\dagger\,(l)}, {c_{B\,\textbf{k}}}^{\dagger\,(l)}\right)$. This does not affect our calculations, presentation of key physics, nor conclusions for this was only an accidental misimplementation during the drafting phase of the paper.
	\item The statement $\Pi^{r} \rightarrow \sum_{j,\,j'}\Pi^{r}_{jj'}$ above Eq. (D1) should be restated as:
	$\Pi^{r} \rightarrow \frac{1}{4}\sum_{j,\,j'}\Pi^{r}_{jj'}$. The division by 4 is to give the sense of averaging over four components of $\Pi^{r}_{jj'}$, when A and B sublattices are indistinguishable. This statement is not used elsewhere in the paper and, therefore, does not affect our calculations, presentation of key physics, nor conclusions.
	
\end{enumerate}

Second, the typographical errors are corrected as below:
\begin{enumerate}
	\item The expressions for the $f(\textbf{k})$ function appearing in Eqs.(A1) and (7) have the wrong signs for $k_{x}$ in the second and third terms. They should be as follows:\\ 
	$f(\textbf{k})=e^{-i\,k_{x}a_{0}} +\,e^{i\,k_{x}a_{0}/2+i\,\sqrt{3}k_{y}a_{0}/2}+
	\,e^{i\,k_{x}a_{0}/2-i\,\sqrt{3}k_{y}a_{0}/2}$.
	\item The expression for $q_{z}$ should have $\lim_{\,\tilde c\rightarrow \infty}$, instead of $\lim_{\,\tilde c\rightarrow 0}$.
	\item In Eq. (B3), the denominators on the right-hand side should be multiplied by $a_{0}$ to make dimensions match for the Laplacian.
	\item In Eq. (D3), the $z'$ and $z$ are taken to be $d^{-}$ in the end ($z'= z= d^{-}$).
\end{enumerate}

Third, we re-clarify some points in discussing the disparity of heat flux ratios between our full RPA model and the local conductivity model: 
\begin{enumerate}
\item We inappropriately adopted the analytical expressions for the plasmon branches ($\omega_{L}$ and $\omega_{H}$ \cite{PhysRevB.92.144307}) valid only for highly doped sheets deviated by mild temperature difference (see Eq. (\ref{revised_wLH}) with $\mu \gg k_{B}T$). The inappropriateness can be obviously seen in the parameters of doping as light as $0.1 eV$ and temperatures as $300K$ and $1000K$ used in our original paper. To remedy this, we derive a more general form:

\begin{equation}\label{revised_wLH}
	\omega_{L/H} = \Bigg[\,Z_{avg}\,\Bigg( 1 \mp \sqrt{1 - (\frac{Z_{1}}{Z_{avg}})(\frac{Z_{2}}{Z_{avg}})(1 - e^{-2qd})}\,\Bigg)\,\Bigg]^{1/2} \qquad,\, q > 0, \tag{E1}
\end{equation}
where the $\omega_{L}$ corresponds to the minus branch and $\omega_{H}$ the positive; $Z_{avg} = (Z_{1}+Z_{2})/2$;\\ $Z_{l} = (e^{2} k_{B}T_{l} / \epsilon_{0} \pi \hbar^{2}) \ln[2\cosh(\mu_{l} / 2 k_{B}T_{l})] \, q$ \,\cite{Ilic_2012}; the $\eta_{1}$ ($ = 0.0033 eV$) was neglected because it is very small compared to the energy spans we have chosen. One can readily verify that this formula goes back to the form $\omega_{L/H} = \omega_{s}\,\sqrt{ 1 \mp e^{-qd}}, \omega_{s} = \sqrt{Z_{avg}}$, when $Z_{1} = Z_{2} = Z_{avg}$ \cite{PhysRevB.92.144307}. One can also show that $\omega_{L} \rightarrow \sqrt{(Z_{1}Z_{2}/Z_{avg})\,qd}$ as $qd \rightarrow 0$. This way of estimating plasmon branches consider the Drude part of conductivity only,
which is fair when the frequency (or $q$) is small and is dominant in wider frequency span when $\mu$ is larger than $k_{B}T$ \cite{Ilic_2012}.

The purpose for drawing the $\omega_{L}$ and $\omega_{H}$ lines using Eq. (\ref{revised_wLH}) is for approximating the orientations of the two branches on the $\omega-q$ plane. The approximation can tell the initial orientations of the plasmon branches in the low frequency regime within the local conductivity model.  

We replace the cyan and green lines in the FIGs. 8 - 9 of our original paper with Eq. (\ref{revised_wLH}) and give the updated plots as follows (the figure indexing follows the same figure indexing in the original paper):

\begin{figure}[htbp]
	
	\subfloat{\modsubfigimg[width=0.475\columnwidth,height=0.365\columnwidth]{\textbf{\textcolor{white}{(a)}}}{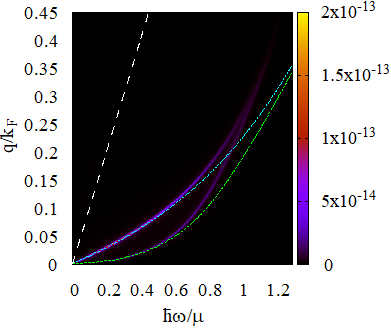}\label{Disp_mine_0.7_10nm}}
	\subfloat{\modsubfigimg[width=0.475\columnwidth,height=0.365\columnwidth]{\textbf{\textcolor{white}{(b)}}}{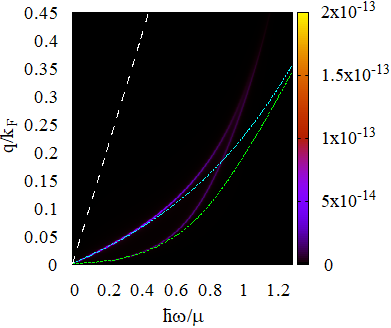}\label{Disp_Ilic_0.7_10nm}}
	\captionsetup{justification=raggedright,
		singlelinecheck=false
	}
	\caption[]{\label{Nonlocal&local_Disp_10nm}\small The updated FIGs. 8 (a) and (b) with the cyan and green dashed lines based on Eq. (\ref{revised_wLH}) replotted; curves standing for Eq. (\ref{revised_wLH}) in the $\mu \gg k_{B}T$ limit are not plotted because the results closely overlap with the cyan and green dashed lines, respectively; the white dashed line marks the $\omega = v_{F}q$ border; other contents are the same as the original paper.}
	
	\subfloat{\modsubfigimg[width=0.475\columnwidth,height=0.365\columnwidth]{\textbf{\textcolor{white}{(a)}}}{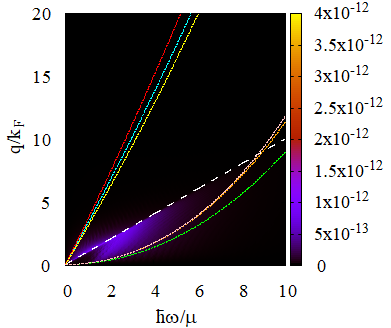}\label{Disp_mine_0.1}}
    \subfloat{\modsubfigimg[width=0.475\columnwidth,height=0.365\columnwidth]{\textbf{\textcolor{white}{(b)}}}{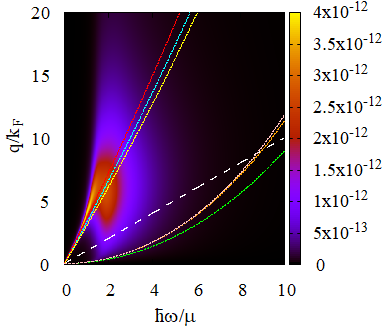}\label{Disp_Ilic_0.1}}\\
    \subfloat{\modsubfigimg[width=0.475\columnwidth,height=0.365\columnwidth]{\textbf{\textcolor{white}{(c)}}}{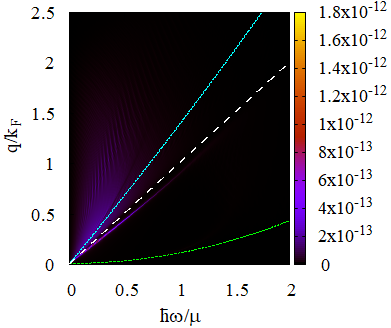}\label{Disp_mine_0.7}}
    \subfloat{\modsubfigimg[width=0.475\columnwidth,height=0.365\columnwidth]{\textbf{\textcolor{white}{(d)}}}{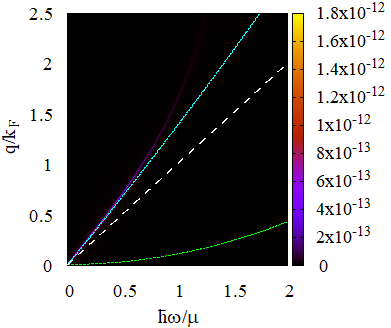}\label{Disp_Ilic_0.7}}
    \captionsetup{justification=raggedright,
	singlelinecheck=false}
    \caption[]{\label{Nonlocal&local_Disp}\small The updated FIGs. 9 (a)-(d) with the cyan and green dashed lines based on Eq. (\ref{revised_wLH}) replotted; the red and orange dashed lines stand respectively for $\omega_{L}$ and $\omega_{H}$ of Eq. (\ref{revised_wLH}) in the $\mu \gg k_{B}T$ limit (these lines are not plotted for (c) and (d) because the results closely overlap with the cyan and green dashed lines, respectively); the yellow and pink dashed lines were mistakenly and respectively taken as the red and orange dashed lines, in the original paper; the white dashed line marks the $\omega = v_{F}q$ border; other contents are the same as the original paper.}
\end{figure}

\item The values of "critical distances" $d_{c}s$ are nearly the same as previously reported \cite{Caroli_PhiG} so we don't change them in the light of Eq. (\ref{revised_wLH}). 

\item We report misplacement of the cyan $\omega_{L}$ lines drawn on FIGs. 9a and 9b in the original paper. The correct lines can be obtained from Eq. (\ref{revised_wLH}) and with the $\mu \gg k_{B}T$ limit applied (see the red, orange, yellow, and pink dashed lines FIGs. \ref{Disp_mine_0.1} and \ref{Disp_Ilic_0.1} in this erratum). 
Also, we report misplacement of figures: FIGs. 8a and 8b in the original paper were mistakenly interchanged due to the ultra-similarity (the local-conductivity curves should bend more slightly inwards toward the $\omega = v_{F}q$ border and appear less blurry at the high-frequency ends).

\item Our previous statement in the original paper: ``The full RPA calculation of ours has rescued the extinction of acoustic plasmon mode under local conductivity approximation by constraining the mode to stay within border of $\omega = v_{F}q$ line." is not well-phrased and confusing, for the full RPA calculation doesn't need to \emph{rescue} its acoustic plasmon mode from invalidating the basic assumption ($\omega > v_{F}q$) that the local conductivity model took. The better statement is: ``The full RPA calculation of ours seems to make the acoustic mode line stay within the border of $\omega = v_{F}q$ line."
\end{enumerate}

Fourth, the lower bound of the characteristic distance scale comparable to the graphene thermal length should be set to $1$ nm (it was $10$ nm in the original paper), corresponding to $\hbar\,v_{F}/0.7 eV \simeq 0.942$ nm ($0.7$ eV is our highest energy parameter). We made this wrong estimation because we did not notice the importance of dividing the estimated wavelengths by $2\pi$ when it no longer indicates a phase of a full cycle of wave in the exponent of $e^{-|q_{z}|d}$. We keep the upper bound ($100$ nm) unchanged, for the characteristic distance scale is now changed to $\hbar\,v_{F}/(k_{B}\times300K) \simeq 25.5$ nm ($k_{B}\times300K$ ($\simeq 26$ meV) is our lowest energy parameter). Thus the range of the characteristic distance scale is now $1-100$ nm. It is checked that the correction only re-adjust our original statement and does influence other parts of the work.

\bibliography{PhiG_NEGF_Erratum}

\end{document}